# An exploratory behavioral and electroencephalographic study of artificial intelligence-assisted learning modes in high school students (June 2026)

**KASHIKA KHURANA [1*], ALLY LIEW [1*]**

[1]Allen High School, Allen, TX, 75013 USA
Corresponding author: K. Khurana (e-mail: kashikakhurana26@gmail.com)

**ABSTRACT** As artificial intelligence (AI) is rapidly integrating into education, concerns have emerged regarding its potential implications on cognitive engagement and problem-solving behavior. However, existing research largely treats AI exposure as a binary condition (AI vs. no-AI), with limited differentiation between interaction modalities and post-exposure effects. This study investigates whether distinct AI interaction modes (Tutor, Collaborator, Solver) influence frontal EEG spectral activity. Electroencephalography (EEG) data and quantified behavioral metrics were recorded from 48 study participants (24 males, 24 females; ages 14–18) across two counterbalanced quizzes in a within-subject design.

Statistical analyses included Friedman tests, repeated-measures ANOVA, paired t-tests, and effect size calculations. Behavioral changes were mathematically analyzed in an observation matrix of three characteristics – Initiation, Processing, and Stress – measured on an ordinal scale. Each mode showed significant differences in all three behavioral measures. Descriptive EEG patterns in AI interaction mode were observed and the possibility of short-term carry over effects of AI were explored. Although the EEG data did not reach statistical significance, the patterns observed across the three AI interaction modes may warrant further investigation. This study provides preliminary behavioral evidence and investigative electrophysiological observations, exploring possible AI-interaction-mode-based differences in neural activity and behavior, while establishing a replicable framework for future human-AI interaction studies.

## I. INTRODUCTION

The rapid development of Large Language Models (LLMs), such as Gemini AI, Claude AI, and ChatGPT, has fundamentally changed the way students learn and solve problems. In educational contexts, artificial intelligence (AI)-driven tools offer opportunities to enhance student engagement and personalize learning to the students' needs. [1] However, the rise of artificial intelligence has also sparked debate regarding its impact on cognitive development in children. Despite the widespread adoption of AI, there remains a limited body of research concerning the neurological effects of these tools, particularly among high school learners. Recent research has begun to explore AI's impact on cognition, but with a focus on higher education. For instance, a 2025 MIT Media Lab project report [2] found that university students relying on LLMs for essay composition exhibited significantly reduced neural connectivity and activation of brain networks compared to those engaged in independent writing. On the other hand, research published in *Frontiers in Education* suggests that AI can optimize learning effectiveness by promoting autonomous and interactive learning environments [3]. Additional studies on design students in Eastern China found that AI-generated content (AIGC) tools were associated with improved creative performance and concentration levels, as measured by electroencephalography (EEG) monitoring [1].

While these prior findings provide a foundational understanding of the implications of AI tools, several aspects remain to be tested.

First, existing studies predominantly focus on university-level students, while leaving the impact on high school students largely unknown. High school students (ages 14-18) are a demographic that use AI as much as university students, but are at a more critical stage of cognitive development, making them a relevant population for studying potential cognitive effects of AI use.

Second, most research treats AI use as a binary condition (use vs. non-use). There is little evidence on whether specific *modes* of AI interaction influence spectral dynamics differently. While the concern that AI use may reduce active learning over time due to decreased cognitive effort during academic tasks [1] poses validity, AI use is not uniform across students. Additionally, current research often isolates single learning styles (e.g., only writing), failing to account for the multifaceted cognitive processes students employ daily, such as math, writing, comprehension, and innovation.

Consequently, it remains difficult to determine whether AI use absolutely diminishes learning-related neural engagement or whether its impact depends on how it is used.

This pilot exploratory study seeks to investigate two primary objectives: (1) whether varied AI interaction modes are associated with differences in behavioral measures and frontal EEG activity, and (2) whether AI exposure may create a short-term effect on oscillatory activity in subsequent independent problem-solving.

To increase ecological validity, this research utilizes a consumer-grade EEG headset (SereniBrain EEG Neurofeedback Headband Model TH21A), providing minimal disruption to a student's typical learning, relative to laboratory-grade EEG systems. This headset may also reveal the potential for a scalable, low-cost type of monitoring in classroom environments.

The implications of this work are significant for both educational policy and software development. By identifying whether different AI interaction modes are associated with distinct behavioral and electrophysiological patterns, this study can inform discussions regarding AI integration into educational curricula and aid engineers to develop artificial intelligence that serves as a cognitive aid rather than a replacement for independent thought.

## II. EXPERIMENTAL AI INTERACTION MODES

Because no common taxonomy exists that strictly defines different AI interaction types, this study creates and utilizes three measurable "modes" for human-AI interaction based on recurring AI-interaction patterns found in educational settings. These modes seek to operationally define the three distinct use cases students often adopt to provide experimentally controllable conditions.

<u>Tutor mode</u>: High learner contribution, low AI guidance.

AI provides Socratic prompting and hints, but is restricted from providing final solutions. This mode requires the student to remain in control and produce the final output independently.

<u>Collaborator mode</u>: Shared contribution between learner and AI.

AI collaborates with the student to brainstorm ideas and complete specific tasks. This represents an interactive cognitive state where executive control fluctuates between the human and the AI model.

<u>Solver mode</u>: Low learner contribution, high AI guidance.

AI provides the solution with no student input. The student may rely fully on the externally generated response.

## III. QUIZZES

Each participant completes two similar difficulty, parallel structure quizzes with three task blocks corresponding to rotating AI interaction modes. Each of the three task blocks test a unique cognitive method (convergent, divergent, and analytical), with one "with-AI" question and one "without-AI" question.

AI modes randomly rotate between task blocks to prevent any order effects.

For example, Participant #1 may have the order Tutor, Collaborator, Solver, Participant #2 may have the order Collaborator, Solver, Tutor, while Participant #3 has the order Solver, Tutor, Collaborator.

Participants were randomly assigned to one of these three orders and researchers ensured that each of the three potential orders had 12 participants each (48 participants total).

Quizzes were designed to assess convergent, analytical, and divergent thinking across math, reading, and open-ended problem-solving tasks. Questions were formatted similarly to questions found on the Scholastic

Assessment Test (SAT) and standardized language tests under the advice of high school math and English teachers.

Quizzes used (Quiz A and Quiz B) were as follows:

**QUIZ A**

Math - Convergent Thinking

1. **Without AI:** A circular garden has a walking path around it that is 2 meters wide. The radius of the garden itself is 8 meters. What is the area of the walking path?

2. **With AI:** In the coordinate plane, a circle has center C at (2,–1), and passes through point A at (6,3). Point B is also on the circle and has coordinates (–2,3).

   What is the area of triangle ABC?

Written Response - Analytical Thinking

1. **Without AI:** Write about the 9/11 Twin Towers attack and its impacts.

2. **With AI**: Write about the COVID pandemic and its impacts.

Innovation - Divergent Thinking

1. **Without AI:** Come up with a creative solution to combat plastic waste.

2. **With AI:** Come up with a creative solution to combat excess water usage.

FIGURE 1. **Quiz A given to all 48 participants. Note that the without-AI questions are before the with-AI questions in all three thinking types (convergent, analytical, and divergent).**

**QUIZ B**

Math - Convergent Thinking

1. **With AI:** In triangle ABC, point D lies on side BC such that BD=3, DC=7. The length of side AB is 10 and the length of side AC is 14. Point E is inside triangle ABC such that:
   - DE⊥BC, and E lies on the angle bisector of ∠A.

   What is the length of DE?

2. **Without AI:** A square has a side length of 8cm. A right triangle is drawn inside the square so that its legs lie along the sides of the square and the hypotenuse of the triangle connects the opposite corners of the square. Find the triangle's area. If the square's side is increased by 20%, what is the new area of the triangle?

Written Response - Analytical Thinking

1. **With AI:** Write about the importance of democracy in society.

2. **Without AI**: Write about the importance of equality in society.

Innovation - Divergent Thinking

1. **With AI:** Describe a global initiative where kids from developing and developed countries work together to learn English.

2. **Without AI:** Describe a global initiative that could help developing and developed countries work together to reduce global warming.

FIGURE 2. **Quiz B given to all 48 participants. Note that the with-AI questions are before the without-AI questions in all three thinking types (convergent, analytical, and divergent).**

## IV. RESEARCH QUESTIONS

The research goals defined above were divided into three components, each seeking to analyze a different aspect of the quizzes.

**Part I**: Do distinct behavioral patterns emerge when students use AI as a Tutor, vs. Collaborator vs. Solver?

**Part II:** Does frontal oscillatory activity differ when AI is used as a Tutor, vs. Collaborator vs. Solver?

**Part III**: Does AI-interaction leave a temporal carryover effect on engagement-related oscillatory activity?

## V. HYPOTHESES:

**Part I**: Participants will demonstrate distinct behavioral cues in each AI mode. Tutor and Collaborator modes are hypothesized to stimulate behavioral patterns related to processing and stress, while Solver mode may promote characteristics of idleness in the participants.

**Part II**: EEG analyses were conducted in an exploratory nature to see if each mode of AI interaction will be associated with differing patterns of frontal EEG oscillatory activity, compared to a no-AI baseline condition and across AI conditions. No confirmatory hypotheses are made regarding directionality due to the exploratory nature.

Frontal activity is operationalized using EEG signals recorded from frontal electrode sites (e.g., Fp1, Fp2, Fpz). Signals will undergo standard preprocessing, including bandpass filtering and artifact rejection prior to spectral analysis.

Power spectral density (PSD) will be estimated using Welch's method, and band power will be computed as relative spectral power (band power normalized to total power across 0.5-40 Hz), in order to reduce skewness.

**Part III**: EEG spectral activity was explored to see if the immediate post-AI task period will exhibit differing short-term frontal oscillatory activity, relative to a pre-exposure baseline. No confirmatory hypotheses are made regarding directionality due to the exploratory nature.

## VI. VARIABLES

Part I & II:
Independent variable:
- AI mode (Tutor, Solver, Collaborator)

Dependent variable:
- I: Qualitative behavioral observations (Initiation, Processing, Stress)
- II: Neurological observations: frontal EEG spectral activity (brain waves recorded: Delta, Theta, Alpha, Beta, Gamma, and the sensorimotor rhythm [SMR])

Part III:
Independent variable:
- Quiz type
  - Quiz A: independent solving first, then with-AI problems
  - Quiz B: with-AI problems first, then independent solving

Dependent variable:
- Changes in frontal EEG spectral activity (brain waves recorded: Delta, Theta, Alpha, Beta, Gamma, and the sensorimotor rhythm)

Constants:
- LLM (ChatGPT)
  - ChatGPT's temporary chat used for all
- Headband (EEG biofeedback headband manufactured by SereniBrain)
  - Sanitized the dry electrodes using isopropyl alcohol-based technical wipes after each use
- Baseline recording time for each quiz (1 min)
- Solving time for each quiz (12 mins)
- Solving time for each question (2 min)
- AI prompts for each mode (See Section VIII)
- 0.5 mm ballpoint pen as writing utensil
- Quizzes (A and B)
- Testing environment (indoors, standardized light, consistent table and chair set up, minimized noise, stable room temperature (75℉)
- Laptop Consistency: M3 MacBook, screen size, resolution, and keyboard used for all participants
- Instructions given
- No caffeine intake in the last 10 hours

Counterbalance variable:
- Order of quizzes given -to control for order effects such as fatigue and reliance (24 participants take Quiz A first and 24 take Quiz B first).

## VII. HUMAN PARTICIPANTS

Composition: 48 high school students (24 males, 24 females; ages 14-18) were recruited for this study.

Participants were also drawn from an academically uniform cohort (weighted GPA of 4.0+, minimum of one AP/IB class) to minimize contextual variability in academic performance that may alter interactions with AI. All participants self-reported having used AI on a daily basis for at least two years.

To reduce potential variability in language-processing backgrounds during text-based interaction tasks, participant selection was composed solely of bilingual students (English and another language both self-reported as first languages, having learned from birth). [See section XIX for a detailed explanation].

Participant demographics are as follows:

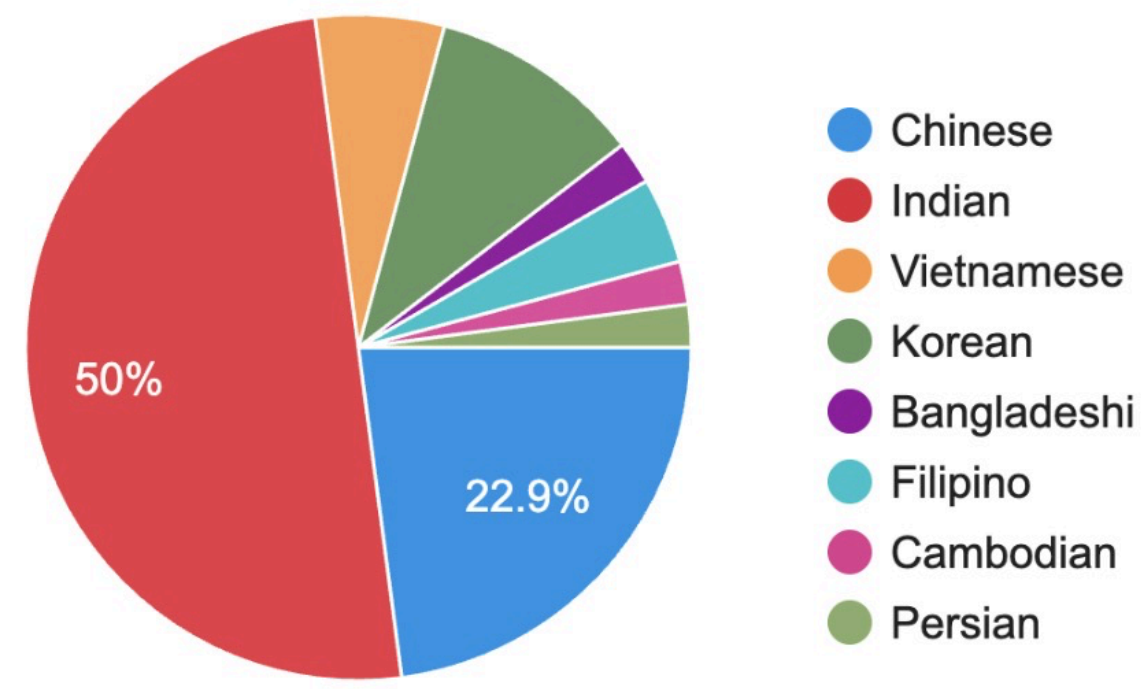

FIGURE 3. **Participant demographic. Ratios represent a sample from the largest public high school in Texas, USA.**

Although ethnicity was not used as a recruitment criterion, all participants in the final volunteer sample self-identified as Asian.

Informed consent was obtained from all participants and their legal guardians (as they were minors). The research plan was approved by the Allen High School Institutional Review Board (IRB) prior to data collection.

## VIII. AI PROMPTS BY MODE

All prompting sessions were executed in temporary ChatGPT sessions (version used being GPT-5 Instant) using the standardized prompt templates below to minimize variations in responses.

TABLE 1. AI PROMPTS BY MODE

| | | |
|---|---|---|
| Tutor | Only steps to solving given. Answer is provided in iterative statements. | You are a tutor helping the student learn how to solve the problem. Do NOT give the final answer.<br>Instead:<br>• explain relevant concepts<br>• ask guiding questions<br>• help the student reason through the steps<br>Your goal is to help the student think through the solution themselves. |
| Collaborator | Possible starting points given, LLM sets broad scope of answer and bullets/guiding questions given. | You are a collaborator working together with the student. Explain the reasoning or approach to solving this problem. No need for step by step. Give the final answer. |
| Solver | Answer fully given. Prompt corresponds to direct output. | You are a solver answering the problem for the student. Provide the full solution. Don't explain how it was obtained. |

## IX. EEG SPECIFICATIONS

1. The device used for experimentation was the SereniBrain EEG Neurofeedback Headband (Model TH21A). This headband was fitted using the International 10-20 system for electrode placement, in which electrodes are placed at either 10% or 20% intervals of the skull's measured distance [4].

   Active biopotential data was collected from two sites:
   - Fp1: Frontopolar 1 (left side of the forehead).
   - Fp2: Frontopolar 2 (right side of the forehead).

   Initial common-mode rejection and differential potential tracking were achieved via a shared central reference electrode at one site:
   - Fpz: Frontopolar midline (center of the forehead).

   The band monitored the frontal lobe region and was placed on the student in accordance with manual instructions (power button aligned with bridge of the nose and lower edge of the band placed immediately above eyebrows) to ensure stable contact over the targeted 10-20 coordinates.

2. Technical Specifications:
   - Sampling Rate: 256 Hz
   - Resolution: 12 to 14 bits per sample
   - Electrode Type: Dry surface biopotential sensors to reduce preparation friction while maintaining continuous structural contact
   - Raw Data Units: Raw voltage fluctuations were recorded in microvolts ($\mu V$).
   - Data Bandwidth: Captures standard EEG frequency bands (delta, theta, alpha, beta, gamma, and SMR) [5]
3. The digital signal processing (DSP) pipeline used involved three primary stages:
   - Artifact Suppression: Signal conditioning includes notch filtering (50/60 Hz) for power-line interference and band-pass filtering (0.5 - 40 Hz) to retain conventional EEG frequency ranges while attenuating background physiological noise [6].
   - Signal Referencing: A common-mode rejection strategy is applied by calculating differential potentials between the frontopolar active sites and the central reference electrode (Fpz) to cancel shared environmental noise.
   - Feature Extraction: Signals are segmented into discrete epochs for Fast Fourier Transform (FFT) analysis. This allowed spectral power estimation across standard frequency bands [7].

To prevent impacts to the pipeline from macro-movements, a real-time contact-quality evaluation was performed via the device's integrated software interface immediately prior to and during experimental trials to confirm baseline data-stream stability.

Because the dry-sensor technology lacks the alternating-current circuitry required to calculate exact laboratory skin-electrode impedance, contact integrity is monitored continuously through automated mathematical tracking of the incoming voltage time-series. The integrated software monitors the stream of data for irregularities, looking for signal saturation, electrical clipping, or abrupt dropouts. Any data exhibiting these unphysiological voltage spikes or a prolonged loss of contact with the skin are removed by the system, removing signal segments with detected artifacts in the final dataset.

## X. METHODOLOGY

1) Data Collection

1. Participants were seated and fitted with a non-invasive EEG headset in accordance with electrical safety standards for EEG [8] and manual instructions. Signal quality was checked prior to data collection. All procedures were conducted under standardized testing conditions as described in Section VI.

2. Prior to seeing the quizzes, a 60-second baseline EEG recording was collected for each participant to control for individual differences in resting neural activity.

3. Participants first completed either Quiz A or Quiz B (see section III) for 12 minutes while researchers noted behavioral observations using a predefined standardized rubric. (Refer to section XIII)

4. This procedure was repeated for a second quiz (12 min, either Quiz B or Quiz A), which reverses the with-AI and without-AI questions to reduce potential order effects and improve internal validity.

2) Data Processing

1. All 96 recordings were offloaded from the SereniBrain data collection application and imported into Kaggle as CSV files. Our time-series data reports values for gamma, beta, alpha, SMR, theta, and delta waves by every 80ms. Our raw exported experimental recordings are categorized by participants and are made to be openly available at the following web URL. (Note that data sharing consent forms collected from participants and their parents). https://www.kaggle.com/datasets/allyliew/participant-data

2. Data cleaning performed, artifacts (eye blinks and head movement contamination) removed, and data z-scored for in-depth analysis. [See Section XI]

3) Behavioral Data Analysis

1. Each participant's behavior scored individually by both researchers.
2. Inter-rater reliability tests performed
3. Behavioral metrics defined by mode. [See Section XIII]

4) EEG Data Analysis

1. Neural activity was later segmented into epochs corresponding to task engagement periods for data analysis.

2. Artifact removal performed on EEG data.

3. Data standardized and baseline z-scored.

4. Using Python libraries [9], brain wave changes, carryover effects, and engagement-related metric differences across quiz questions and AI modes were analyzed.

5. Standard EEG waveforms were identified according to established principles [10].

Alpha waves (8–12 Hz) were monitored for relaxation [11], while Beta waves (13–30 Hz) served as the primary indicator of active focus [12]. Theta (4-8 Hz) activity was analyzed to assess internal focus [13]. Gamma (31-40 Hz) was monitored for high-level executive processing [14]. Additionally, cross-frequency coupling between Beta and Gamma bands was utilized to assess functional neural integration [15].

5) Statistical Analysis

1. Behavioral measures (Initiation, Processing, Stress) were designated as the primary endpoints and evaluated for differences across AI interaction modes. Statistical tests included using a Friedman test, followed by Wilcoxon signed-rank post-hoc comparisons with Holm–Bonferroni correction.

2. EEG spectral measures were used as secondary exploratory outcomes. A one-way repeated-measures ANOVA was used to assess overall differences across conditions. EEG results were not subjected to post-hoc inference due to a lack of significant omnibus effects. Accordingly,

EEG findings are reported in descriptive form without correction-based inference.

3. Statistical analysis was performed using the statsmodels (v0.14.0) and scipy (v1.11.0) libraries in Python. Repeated Measures ANOVA (AnovaRM) was utilized to account for the within-subject design [see section XI for in depth information].

# XI. DATA PROCESSING

## A. ARTIFACT REMOVAL

Since EEG recordings are susceptible to substantial physiological and environmental noise, the data was first cleaned for artifacts (such as muscle movements and blinks) that could cause large spikes in oscillatory dynamics [16]. To preserve the reproducibility and efficiency required for classroom settings, a targeted interquartile range (IQR) artifact rejection algorithm was chosen over computationally intensive blind source separation techniques (e.g., ICA).

$$IQR = Q_3 - Q_1 \quad (1)$$

$$Lower\ Bound = Q_1 - 1.5(IQR) \quad (2)$$

$$Upper\ Bound = Q_3 + 1.5(IQR) \quad (3)$$

By using the interquartile range and the upper and lower bounds found, any data that exceeded those ranges could be removed. This process controlled for random high spikes in data resulting from artifacts [5].

Additionally, the raw voltages were smoothed and averaged over a larger window with a simple moving average equation (SMA) – a common preprocessing technique used to reduce high-frequency signal noise and improve signal stability prior to downstream analysis.

$$SMA_t = \frac{1}{W}\sum_{i=0}^{W-1} x_{t-i} \quad (4)$$

The simple moving average (SMA) was taken over a window (W) size of five rows to balance noise reduction with signal preservation. This prevents notable temporal patterns from being excessively smoothed.

## B. BASELINE NORMALIZATION

Baseline z-score normalization was performed to scale each participant's data to their own baseline-resting cognitive state (obtained from the one minute baseline recording). This allowed for accurate cross-participant comparisons by looking at average standard deviations from the mean instead of raw values.

$$z = \frac{x - \mu_{base}}{\sigma_{base}} \quad (5)$$

The z value was calculated by taking the raw EEG voltage at any given time (x) and subtracting it from the participant's average resting voltage ($\mu_{base}$) and then dividing the result by the standard deviation of the resting voltage ($\sigma_{base}$).

## C. LEVENE'S TEST

Levene's test was performed to ensure homogeneity of variances across all AI modes.

$$W = \frac{(N-k)}{(k-1)} \frac{\sum_{i=1}^{k} N_i(\bar{Z}_{i.} - \bar{Z}_{..})^2}{\sum_{i=1}^{k}\sum_{j=1}^{N_i} (Z_{ij} - \bar{Z}_{i.})^2} \quad (6)$$

The raw data was transformed by calculating the absolute deviation ($Z_{ij}$) for each data point, representing the distance between a value and its respective group's median. These deviations were then utilized as a new dataset to conduct a one-way ANOVA. The variance between the four AI conditions (k) – calculated using the group sizes ( $N_i$), group mean deviations ($\bar{Z}_{i.}$), and the grand mean deviation ($\bar{Z}_{..}$) – was compared against the variance within the groups. A resulting W statistic with a non-significant p-value indicated that the spread of data was uniform across the conditions, confirming the assumption of homogeneity of variance.

## D. ONE-WAY REPEATED MEASURES ANOVA

A one-way repeated measures ANOVA was conducted to evaluate whether statistically significant differences in neural activity existed across the three AI interaction modes (Tutor, Collaborator, Solver) within the same participants.

$$F = \frac{MS_{Treatment}}{MS_{Error}} \quad (7)$$

$$SS_{Total} = SS_{Treatment} + SS_{Subject} + SS_{Error} \quad (8)$$

The F-statistic represents the ratio of variance between the experimental manipulation and the unexplained within-subject variability.

Total variability ($SS_{Total}$) was partitioned to reflect differences between AI modes, individual baseline variance, and unexplained residual variation. This method of isolating within-subject changes increases statistical accuracy as brain wave activity becomes independent of inter-individual variability.

### E. PAIRED COHEN'S D

$$d_z = \frac{\bar{x}_{diff}}{s_{diff}} \quad (9)$$

For within-subject comparisons (AI vs. no-AI), paired Cohen's d was calculated. $\bar{x}_{diff}$ represents the average difference between conditions for each participant, and $s_{diff}$ represents the standard deviation of those differences. This approach accounts for repeated measurements obtained from the same individuals across conditions, making it more appropriate than independent-sample effect size calculations for within-subject experimental designs. Effect sizes (d) were interpreted using conventional thresholds, where d ≈ 0.2 indicates a small effect, d ≈ 0.5 a medium effect, and d ≈ 0.8 a large effect. In addition, 95% confidence intervals were considered to evaluate the precision and stability of the estimated effects. In this study, paired Cohen's d was used to quantify the magnitude of AI-related changes in neural activity within individuals, providing a complementary measure of practical significance beyond p-values alone [17].

## XII. STATISTICAL TEST RESULTS

TABLE 2. SIGNIFICANCE TESTING

| Test | Result | Impact |
|---|---|---|
| Paired t-test (AI vs. no-AI) | p-value = 0.823 > 0.05<br>Cohen's d = -0.037 | No significant differences were observed between the with-AI and no-AI conditions, as a negative effect size emerged. |
| One-Way Repeated Measures ANOVA | p-values (across all six bands*) > 0.29 | Stable p-value failed to reject the null hypothesis. This means no statistically significant differences observed across AI modes. |
| Paired t-test (Quiz A vs. Quiz B) | p-values (across all six bands**) > 0.20<br>Cohen's d = 0.11 | Non-significant variance between the results potentially due to the parallel structuring of quizzes. |
| Levene's Test (for variance stability) | p-values (across all six bands***) > 0.53 | Variance homogeneity assumption satisfied. |

**EXTENDED VALUES FOR EACH BRAIN WAVE BAND**
(corresponds to asterisks on Table 1)

*One-way repeated measures ANOVA band values

| | | | |
|---|---|---|---|
| Gamma | F = 0.85, | p = 0.4688 | [ns] |
| Beta: | F = 1.26, | p = 0.2911 | [ns] |
| Alpha: | F = 0.63, | p = 0.5953 | [ns] |
| Theta: | F = 0.70, | p = 0.5523 | [ns] |
| Delta: | F = 1.00, | p = 0.3942 | [ns] |
| SMR: | F = 0.98, | p = 0.4037 | [ns] |

**Paired t-test band values

| | | | |
|---|---|---|---|
| Gamma: | t = -1.30, [ns] | p = 0.2007 | d = -0.19 |
| Beta: | t = 0.75, [ns] | p = 0.4597 | d = 0.11 |
| Alpha: | t = 0.52, [ns] | p = 0.6082 | d = 0.08 |
| Theta: | t = 0.09, [ns] | p = 0.9289 | d = 0.01 |
| Delta: | t = 1.12, [ns] | p = 0.2704 | d = 0.16 |
| SMR: | t = -1.01, [ns] | p = 0.3173 | d = -0.15 |

***Levene's test band values

| | | |
|---|---|---|
| Gamma: | p = 0.7071 | [Stable (Equal Variance)] |
| Beta: | p = 0.5269 | [Stable (Equal Variance)] |
| Alpha: | p = 0.6153 | [Stable (Equal Variance)] |
| Theta: | p = 0.6140 | [Stable (Equal Variance)] |
| Delta: | p = 0.8158 | [Stable (Equal Variance)] |
| SMR: | p = 0.7528 | [Stable (Equal Variance)] |

## XIII. BEHAVIORAL RESULTS

Compared to the EEG results, behavioral tests yielded highly significant differences between the three AI interaction modes.

A structured, blinded behavioral observation matrix was deployed to capture real-time human-factors dynamics. Behavioral execution was quantified across three operationalized dimensions using discrete behavioral metrics. These metrics were calculated using only the observations from the with-AI questions.

1. Initiation: Indication of individual processing before AI input based on duration in seconds between viewing the quiz question and beginning to use the AI model.
2. Processing: Indication of thinking based on the number of pauses in between writing/solving.

3. Stress: Indication of stress based on the number of distinct adjustments associated with high overload or frustration.

To ensure the objectivity of the behavioral observations and eliminate bias, researchers scored each participant independently of each other. The real-time frontal EEG spectral data streams were also not viewed during behavioral analysis to mitigate potential biases.

Students were each given a score from zero to two (with zero equaling less of the behavior and two equaling more of the behavior as observed)

These scores were assigned based on the following guidelines:

Initiation:

- 0 = Immediately uses AI (no attempt to solve independently)
- 1 = Brief attempt (<3 sec of individual processing)
- 2 = Clear attempt (>3 sec writes/thinks before AI)

Processing:

- 0 = No pauses while working
- 1 = Some pauses while working (1–3 times)
- 2 = Frequent pauses while working (3+ times)

Stress:
Participants were scored based on the number of times these indicators of stress were observed [18]:

1. Fidgeting: Tapping foot, drumming fingers, tapping a pencil, or shifting in seat
2. Sighing: observed stress reducing behavior
3. Micro-movements: Picking at skin around fingers, lip chewing, hair twirling, or teeth grinding/ jaw clenching.

Stress score:

- 0 = No signs of stress
- 1 = 1–3 signs of stress
- 2 = 3+ indicators of stress

Thresholds were selected a priori to standardize observational coding and maintain consistency across raters. To ensure statistical reliability, behavioral timelines were independently scored by two researchers (R1 and R2) according to the above standardized rubric, yielding high inter-rater consensus before being subjected to non-parametric Friedman and Wilcoxon signed-rank testing.

**Inter-Rater Reliability Analysis**: Inter-rater reliability (IRR) was quantified across all 144 behavioral scores (48 participants across three distinct operational modes), resulting in $N = 432$ paired metric observations across the three evaluation dimensions (Initiation, Processing, and Stress). The macro-level analysis demonstrated exceptional consensus between the two researchers, achieving an overall percent agreement of 96.30% (416 out of 432 exact matches) and an overall Cohen's kappa coefficient (κ) of 0.9422, which denotes near-perfect agreement under standard benchmarking criteria. High agreement likely reflects the use of behaviorally explicit scoring criteria consisting of counting.

The mathematical formulations utilized for verification are defined as follows:

F. ***PERCENT AGREEMENT:*** The proportion of times both evaluators gave the same score for the same metric.

$$P_o = \frac{\sum_{i=1}^{M} I(R_{1i} = R_{2i})}{N} \tag{10}$$

where $I$ is an indicator function equal to 1 if the ratings match and 0 otherwise, and $N$ represents the total number of paired observations.

G. ***COHEN'S KAPPA:*** To control for agreements occurring strictly by chance, the kappa statistic was computed via:

$$\kappa = \frac{P_o - P_e}{1 - P_e} \tag{11}$$

where $P_e$ represents the hypothetical probability of chance agreement, derived from the marginal totals of the confusion matrix:

$$P_e = \sum_{k=0}^{K} P_{1k} P_{2k} \tag{12}$$

$P_{1k}$ and $P_{2k}$ are the independent marginal probabilities of assigning score category $K$ for R1 and R2, respectively.

Given the ordinal structure of the rubric, weighted Cohen's kappa coefficients were additionally computed to account for disagreement severity between adjacent scoring categories. Granular inter-rater reliability metrics were extracted for each behavioral metric:

- Initiation: 96.53% agreement, $K = 0.9421$, weighted $K = 0.9703$

- Processing: 93.75% agreement, $K$ = 0.9060, weighted $K$ = 0.8671
- Stress: 98.61% agreement, $K$ = 0.9774, weighted $K$ = .9880

These statistically high (almost 1) indices confirm that the standardized behavioral rubric minimized subjectivity and provided a robust foundation for subsequent non-parametric inference testing.

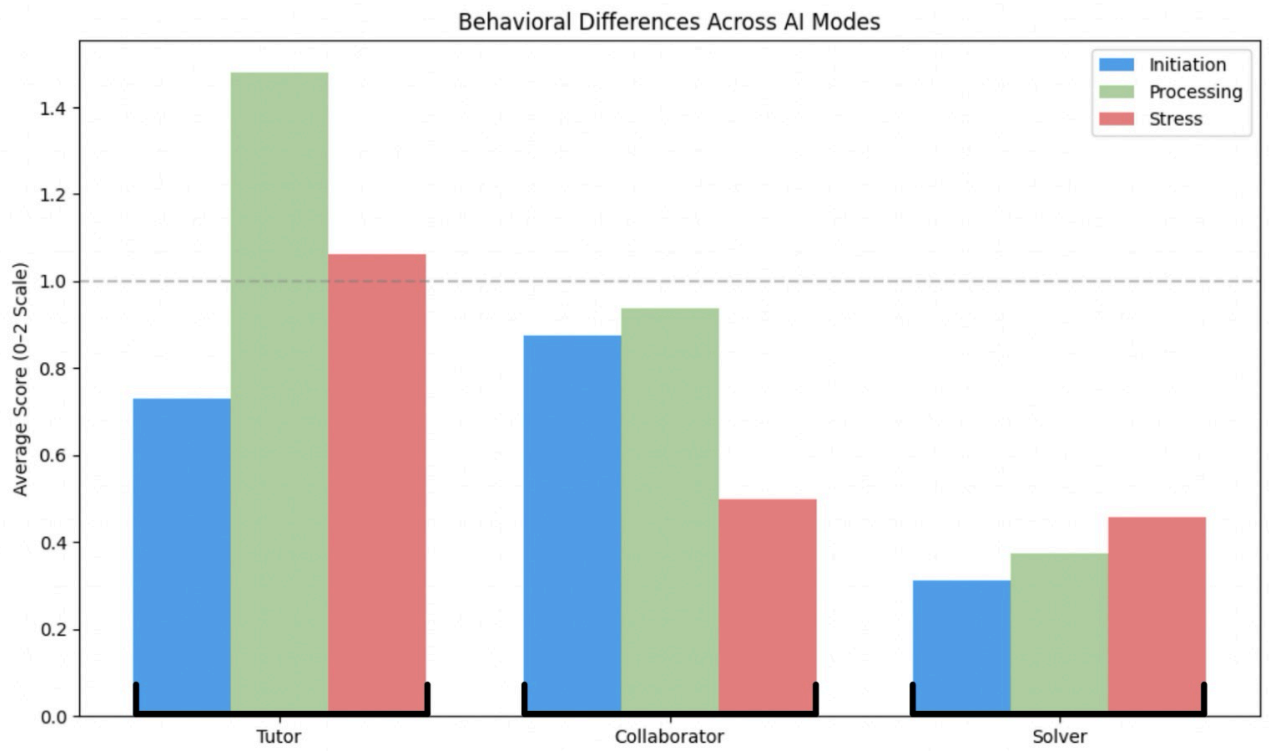

FIGURE 4. **After ensuring high inter-rater reliability, behavioral averages across all three AI modes across three parameters (Initiation, Processing, and Stress) were plotted. See below for significance across groups. Scores can be accessed via this Kaggle dataset:** https://www.kaggle.com/datasets/kashikakhurana/ally-liew-r1-and-kashika-khurana-r2-behavioral-data

To evaluate the differences in Initiation, Processing, and stress, a non-parametric Friedman test was conducted separately for each behavioral dimension. Because the Initiation, Processing, and Stress scores are ordinal, and our study uses a repeated-measures design (where the same participants test all three interaction modes), parametric assumptions of normality are not met. Therefore, the Friedman test, a non-parametric alternative to the one-way repeated measures ANOVA, was implemented [19].

The Friedman test statistic is calculated using the following formula:

$$X^2_F(2) = \frac{12}{Nk(k+1)}\sum_{j=1}^{k} R^2_j - 3N(k+1) \quad (13)$$

where $N$ is the number of participants ($N$ = 48), $k$ is the number of conditions ($k$ = 3), and $R_j$ represents the sum of ranks for each condition $j$. To quantify the magnitude of the overall treatment effect, Kendall's $W$ was calculated as an effect size, where values closer to 1 indicate perfect agreement in rankings across participants.

Specifically:
$X^2$ = chi-square statistic
$F$ = Friedman version of chi-square
(2) = degrees of freedom, because:

$$df = k - 1 = 3 - 1 = 2 \quad (14)$$

where $k$ = 3 conditions.

The analysis revealed highly significant differences across modes for Initiation ($X^2_F(2)$ = 12.88, p = 0.0016, Kendall's $W$ = 0.134), Processing ($X^2_F(2)$ = 38.57, p < 0.0001, Kendall's $W$ = 0.402), and Stress ($X^2_F(2)$ = 21.80, p < 0.0001, Kendall's $W$ = 0.227). These findings indicate that the selected AI interaction paradigm significantly influenced participant behavior patterns.

Because significant Friedman test results indicate only global differences without identifying which specific modes differ from one another, post-hoc pairwise comparisons were conducted using the Wilcoxon signed-rank test [20]. Given the moderate sample size ($N$ = 48 > 20), the normal approximation ($Z$-score) of the Wilcoxon statistic was utilized for effect size estimation. Holm-Bonferroni adjustment was additionally applied to control the family-wise error rate. Reported p-values reflect Holm-Bonferroni-adjusted significance testing.

Post-hoc pairwise comparisons using the Wilcoxon signed-rank test were calculated using the following equation:

$$Z = \frac{T - \frac{N_r(N_r+1)}{4}}{\sqrt{\frac{N_r(N_r+1)(N_r+2)}{24} - \frac{\Sigma(t_i^3 - t_i)}{48}}} \quad (15)$$

where T represents the sum of signed ranks, $N_r$ is the number of non-zero differences, and $t_i$ denotes the size of tied rank groups. To quantify the magnitude of pairwise behavioral difference, effect size ($r$) is calculated using:

$$r = \frac{Z}{\sqrt{N}} \quad (16)$$

where $Z$ = normalized test statistic from the Wilcoxon signed-rank test, and $N$ = number of paired observations.

Initiation showed no significant difference when comparing Tutor and Collaborator modes ($Z$ = -0.956, p = 0.439, r = 0.138), suggesting these two modes trigger similar styles of Initiation prior to AI utilization. However, significant differences were identified between Tutor and Solver ($Z$ = -2.818, p = 0.0048, r = 0.407) and between Collaborator and Solver modes ($Z$ = -3.836, p < 0.001, r = 0.554), showcasing substantially reduced independent

Initiation during Solver interactions.

Secondly, Processing behavior differed significantly across all pairwise comparisons. Tutor mode produced significantly greater Processing activity than Collaborator mode ($Z$ = -3.374, $p < 0.001$, r = 0.487) and Solver mode ( $Z$ = -4.306, $p < 0.0001$, r = 0.622). Collaborator mode also demonstrated significantly greater Processing behavior than Solver mode ($Z$ = -3.793, $p < 0.001$, r = 0.547).

Finally, Stress levels also varied significantly between modes. Tutor mode demonstrated significantly higher stress-related behaviors than Collaborator mode ($Z$ = -3.909, $p < 0.001$, r = 0.564) and Solver mode ($Z$ = -3.113, p = 0.0019, r = 0.449). No statistically significant difference was identified between Collaborator and Solver modes ($Z$ = -0.215, p = 0.829, r = 0.031). It is important to note that these Stress indicators represent observational proxies of cognitive overload rather than direct physiological measurements of stress.

## XIV. EXPLORATORY EEG RESULTS

The behavioral results from above outline the strongest findings from our study. The EEG results below provide exploratory trends that aim to motivate investigation into the causes behind the patterns discovered. However, because of the non-significance of the data and low channel set up, they should be seen as preliminary trends without inferential interpretation.

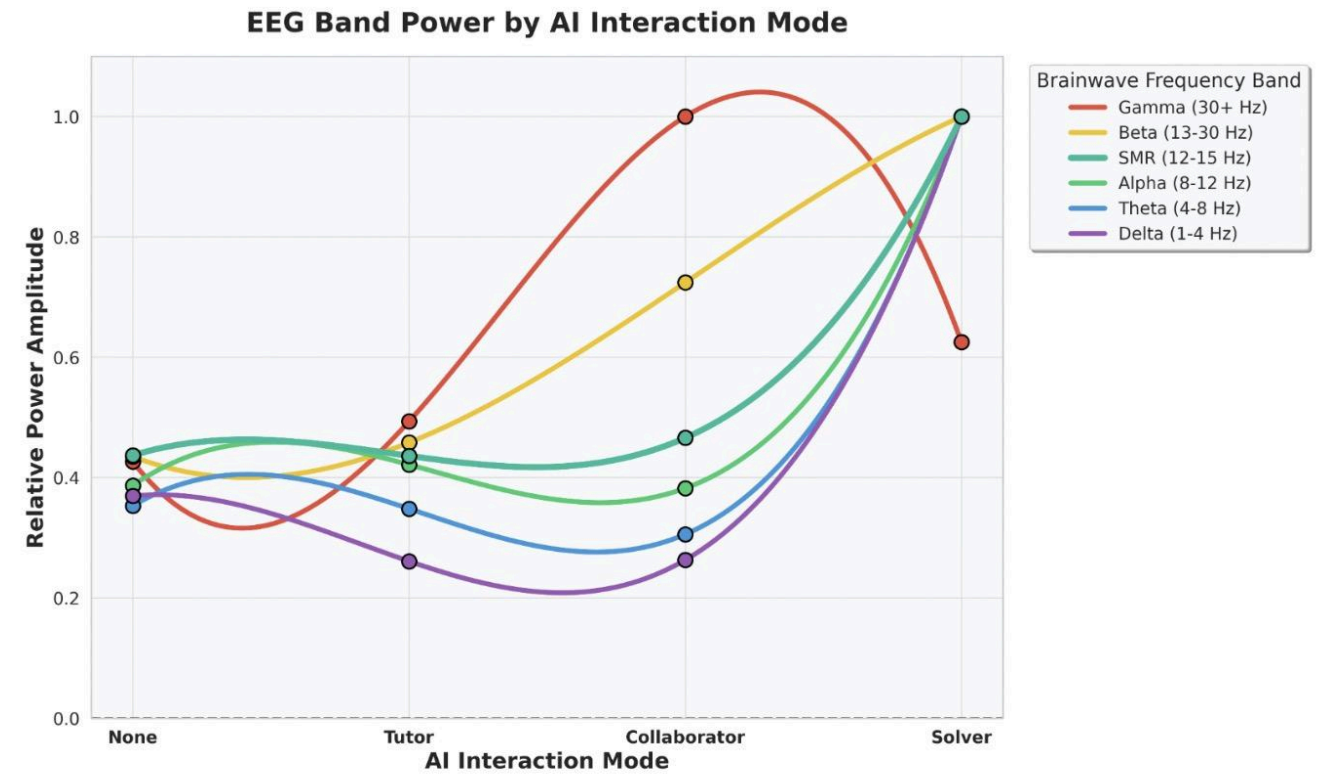

FIGURE 5. **Master graph of all six EEG bands (gamma, beta, SMR, alpha, theta, delta) across no-AI, Tutor mode, Collaborator mode, and Solver mode.**

An EEG measures electrophysiological differences using established band values seen in Figure 5. Figure 5 offers a descriptive idea into how AI interaction modes may affect oscillatory activity differently. Conclusions were found using an exploratory visual inspection as outlined below.

When comparing each AI mode with the control or balanced cognitive state of no-AI (None), Tutor mode displayed the most consistent oscillatory dynamics with the no-AI condition.

In this study, gamma and beta increases were recorded in Collaborator mode. Gamma and beta waves can be associated with active processing and focus, but the rise in gamma and beta through Collaborator mode is not necessarily an indication of this cognitive reaction due to the limited abilities of the headset.

Finally, all waves were observed to have risen except for gamma in Solver mode. Increases in alpha, theta, and delta activity alongside reduced gamma activity are theoretically consistent with patterns of cognitive offloading hypotheses as reported in prior literature [21]. However, whether the pattern in Solver mode represents cognitive offloading or not cannot be determined from this study alone.

Although these inferential conclusions did not reveal significant differences between modes, the descriptive oscillatory trends suggest potential mode-dependent processing variations warranting further investigation.

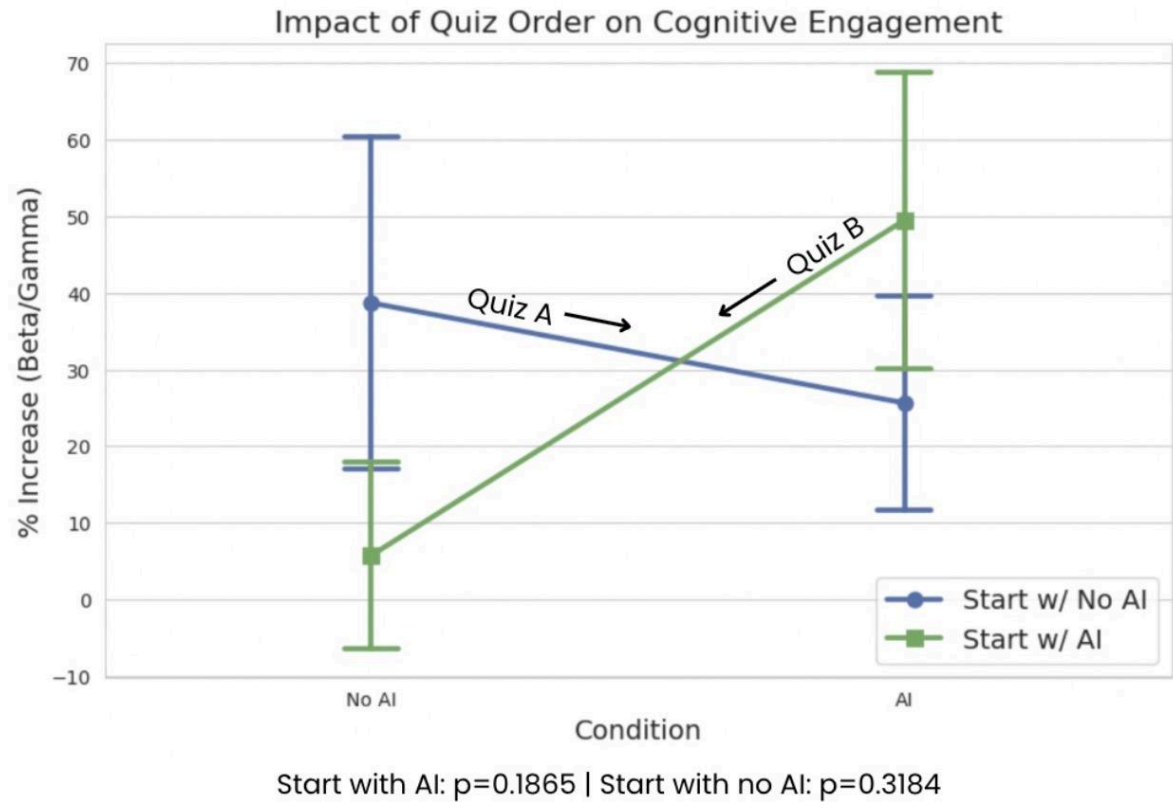

FIGURE 6. **Residual effects of AI across quiz A and quiz B**

When participants completed Quiz A first, a slight decrease in beta and gamma from the initial no-AI condition to the later with-AI condition was observed, but not enough to produce a significant change in engagement-related EEG metrics.

This trend did become more prominent in Quiz B, where completing the AI question first displayed descriptively lower beta and gamma activity when students worked without AI afterwards. It cannot be concluded from this study that these trends reflect a carryover effect from AI due to the non-significance of the data, but it does reveal a topic worth researching further.

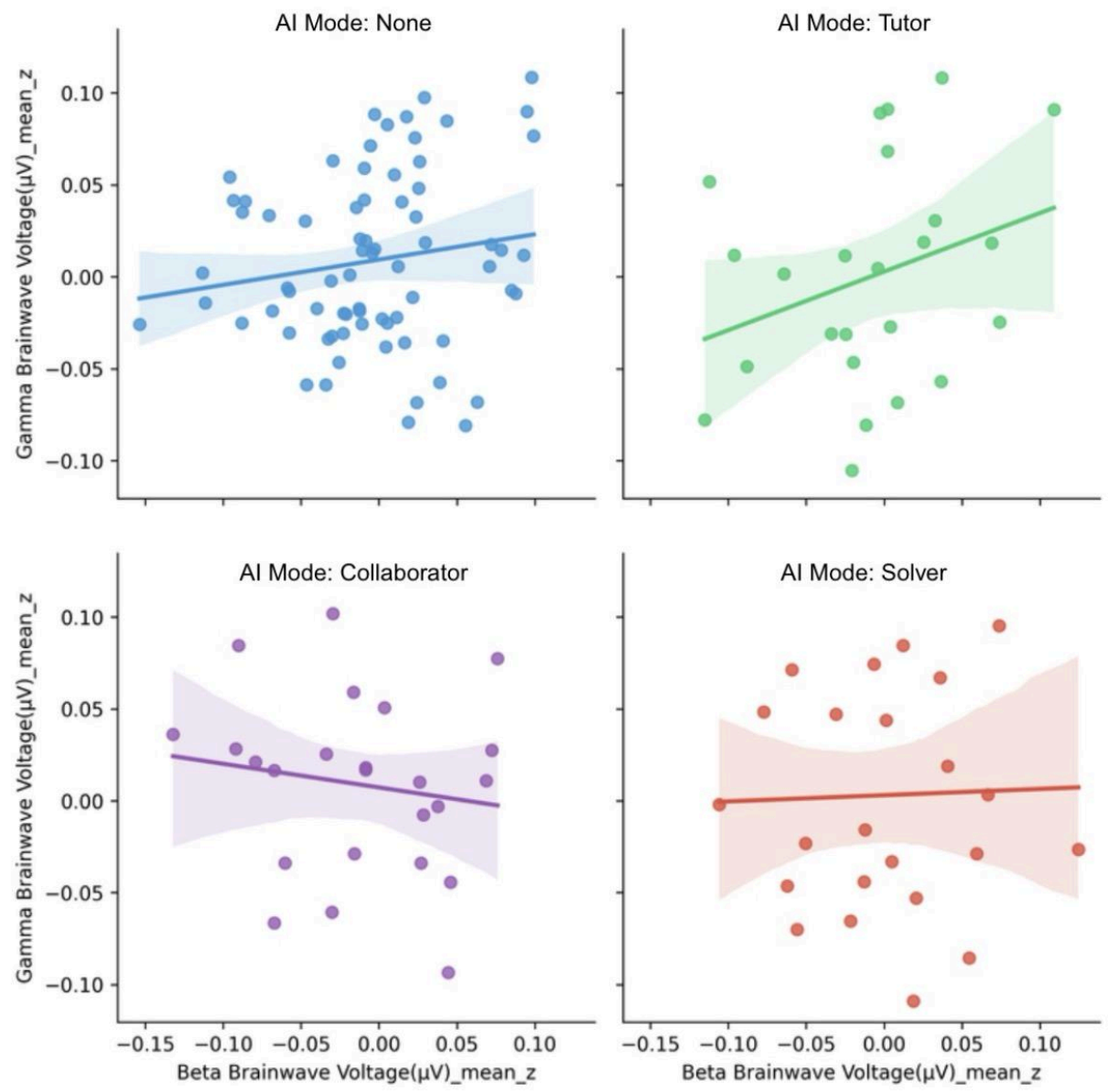

FIGURE 7. **Beta-gamma coupling across none/no-AI, Tutor mode, Collaborator mode, and Solver mode. P-value (no-AI): 0.177, p-value (Tutor): 0.1553, p-value (Collaborator): 0.4787, p-value (Solver): 0.8917**

Prior literature has associated beta-gamma coupling with information integration and executive processing [15]. Although the beta-gamma relationships in this experiment differed descriptively across conditions, none of these correlations reached statistical significance ($p > 0.15$). With these current, non-significant results, only descriptive patterns can be interpreted to produce exploratory rather than confirmatory conclusions.

The no-AI mode served as a baseline model of spectral dynamics where both beta and gamma waves rise together.

Although not numerically significant, a descriptively stronger positive beta-gamma relationship was observed in Tutor mode. A positive correlation between beta and gamma can indicate greater coordination between high-frequency oscillatory activity, but it cannot be determined from this study if greater coordination is what is actually occurring.

An opposite pattern where beta and gamma were negatively correlated was measured in Collaborator mode. Due to the non-significance of the data, the possibility of a shift in the brain's information-processing dynamics cannot be confirmed but may spark future investigation.

Solver mode displayed a flat line, where beta increases but gamma does not increase. This pattern is typically associated with weaker beta-gamma coupling and could warrant further investigation. However, it cannot be concluded that this represents cognitive offloading.

The beta-gamma plots reveal preliminary correlation patterns that may provide useful directions for future research on how AI interaction modes relate to attention and information integration. They suggest that cross-frequency relationships may be worth examining in larger studies with higher-density EEG recordings.

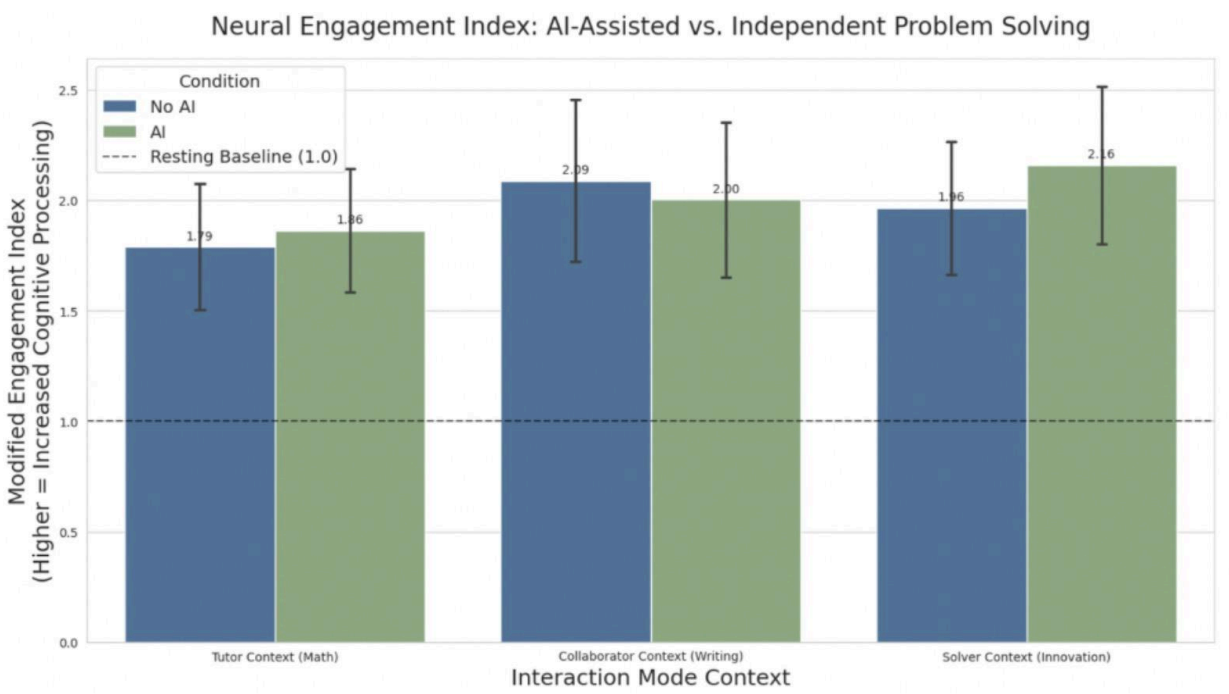

FIGURE 8. **Engagement index values between Quiz A and Quiz B across all three modes: Tutor Context (Math): p-value: 0.8555 | Collaborator Context (Writing): p-value: 0.8649 | Solver Context (Innovation): p-value: 0.6792**

"Engagement" was measured and quantified using an engagement index as an ad hoc exploratory metric.

The engagement index [22] was calculated and averaged for all 48 participants using the following equation:

$$Engagement\ Index = \frac{beta + gamma}{alpha + theta} \tag{17}$$

As the numerator (waves associated with active processing) increases, engagement index increases, and as the denominator (waves associated with idleness and daydreaming) increases, engagement index decreases.

While each mode reflected varying trends in individual brainwaves, overall engagement-related EEG metrics showed no significant differences across all three modes (Tutor, Solver, Collaborator) and between AI vs. no-AI.

## XV. DISCUSSION

This study should be interpreted as a pilot exploratory investigation rather than a definitive test of neural mechanisms. The behavioral data provides the strongest evidence right now, while the use of a consumer-grade EEG headset and absence of statistically significant spectral differences limit causal interpretation. Consequently, the electrophysiological findings are best viewed as descriptive trends that complement the behavioral results and serve as a way to inform hypotheses for future studies using higher-density neurophysiological recordings.

The behavioral evidence suggests that AI interaction types can produce statistically significant differences in behavior depending on the mode. However, Tutor and Collaborator mode demonstrated similar Initiation behaviors. For Processing and Stress indications, Tutor mode revealed the most Processing behavior but also the most Stress. Solver mode promoted the least Stress and the lowest Processing and Initiation.

The lack of statistical significance in EEG data ($p > .29$) suggests that AI interaction mode does not create massive drops or increases in cognitive activity. However, the non-significant EEG results should be interpreted cautiously. Rather than proving that AI interaction has no neurological effect, the current findings indicate that this study did not detect large group-level changes in frontal EEG spectral power under the present experimental conditions. Future work using higher-density EEG, improved artifact control, and larger samples would be needed to evaluate more subtle neural effects.

While the headset used presents a limitation in the data collected, it also highlights the value of this work as a low-cost, thus scalable study able to be integrated into school environments without major disruption to a student's normal work setting. By providing this early framework, the experimental method can be replicated in various educational settings.

EEG data revealed descriptively different activity corresponding with each AI mode. These visual data points could inspire further investigation into the mechanisms behind these responses. Researchers could even explore how different interaction types could potentially lead to specific effects like cognitive offloading and cognitive load [21][23].

This study also introduces a new investigation into AI's short-term carryover effect, which previously has not been researched extensively. Descriptively lower levels of beta and gamma were observed in no-AI conditions. The possibility of AI's association with increased reliance on an external source should be explored with future large-scale experiments, but currently reveals no significant effect.

The engagement index findings displayed no significant difference in "engagement" (measured based on the ratio between beta + gamma brainwaves and alpha + theta brainwaves) regardless of the task. The metric used to calculate "engagement" may be oversimplified but was utilized as part of this exploratory study. Future studies could explore this further, as well as if the brain engages the same amount regardless of AI use but shifts to a different cognitive strategy.

## XVI. CONSTRAINTS

1. As participants moved to the later questions, a potential mental fatigue (characterized by descriptive increases in delta activity) emerged. As this may have affected the accuracy of our results, future studies should test each AI mode individually from each other by providing a 1-minute break between each type of task block.

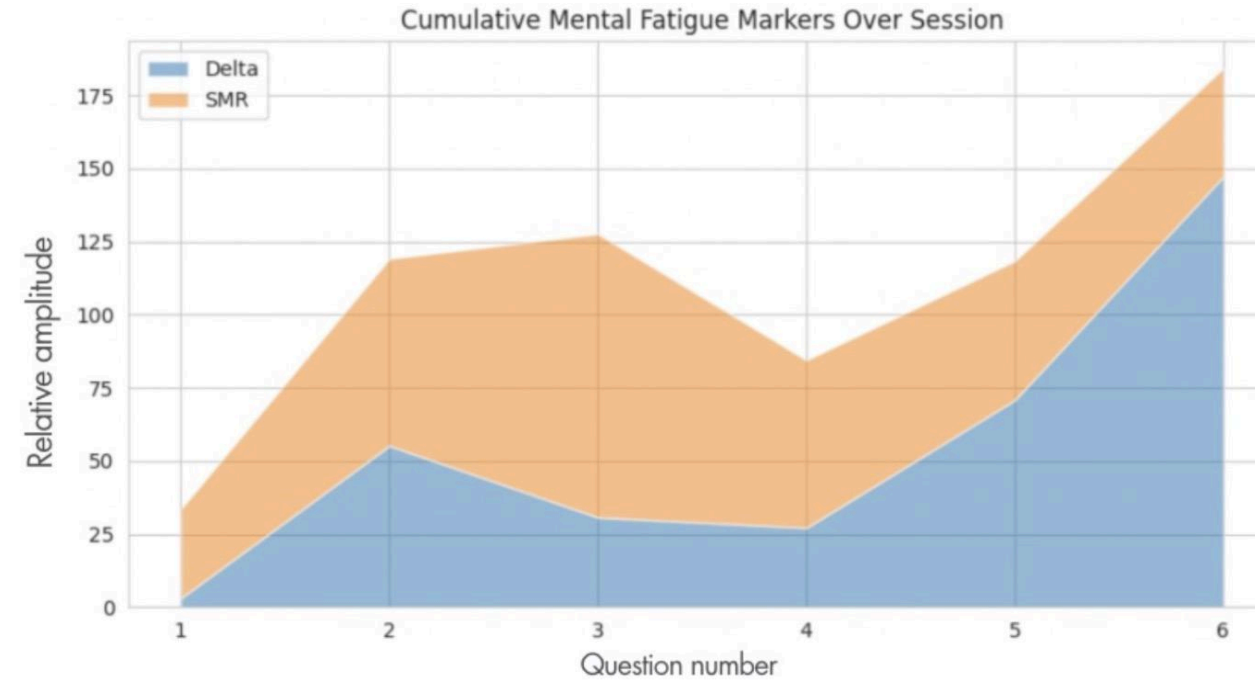

FIGURE 9. **Delta-oscillations between quiz questions**

2. Each person had varying focus levels in the quiz, so variations between participants had a greater effect on the overall data. We controlled for this by having each participant take the quiz two times (Quiz A and Quiz B), but for future studies, it would be advisable to test a larger number of participants.

3. The EEG headset used had limited spatial coverage, which may have affected the accuracy of our data due to its inability to pick up strong oscillations from other regions of the brain beyond the front. Additionally, it was susceptible to contamination without a high-level filter to account for noise. To reduce the amount of artifacts in our data, we used simple data processing techniques such as the interquartile reduction method and a simple moving average, but were unable to perform targeted reduction techniques that require more equipment. While neural differences across modes had a p-value $> 0.29$, descriptive patterns emerged. However, in the future, it is advisable for studies to use a full-coverage spatial headset and accordingly, higher levels of data processing in order to ensure full validity of results.

4. Lastly, we cannot interpret the gamma increase as definitive evidence of higher-level cognitive processing. While gamma-band oscillations are frequently associated with attention and information integration, their interpretation remains largely context-dependent [9]. One such potential confounding variable is a transcription-related motor artifact effect resulting from the gamma wave's sensitivity to active, high-precision physical execution (fast-paced writing) combined with sensory feedback. Because elevated gamma-band oscillations (31–40 Hz) are highly sensitive to micro-muscular artifacts, rather than reflecting purely internalized comprehension or cognitive enhancement, gamma increases may partially reflect sensorimotor integration demands (e.g., head moving, pencil strokes). Future iterations of this experiment should decouple processing from active writing phases.

## XVII. NEW IMPLICATIONS ON AI BRAIN RESEARCH

To our knowledge, this is one of the first studies to experimentally test the unique modes of Tutor, Solver, and Collaborator when analyzing how AI use affects brain waves in teenagers. By breaking down the previously generalized "AI interaction" into three distinct use cases, we were able to find preliminary evidence that AI modality may influence short-term cognitive engagement patterns in high school learners.

For the purpose of this relatively new concept, unique quizzes were created to be used in the context of high school students. In both Quiz A and Quiz B, the math, writing, and innovation questions developed were based on standardized high school exams like the Scholastic Assessment Test (SAT).

We were able to find behavioral cues that participants were exhibiting across each mode, such as:

- How many seconds participants took to try the question independently prior to AI help as a Tutor vs. Solver vs. Collaborator
- How many times participants paused to process the AI's response
- How many times the students showed signs of stress when evaluating AI's response (identified three markers of common behavior when using AI) [see Figure 4]

By implementing our unique parallel quiz structure across all 48 participants (Quiz A and Quiz B), our study was also the first (to our knowledge) to explore carryover effects of AI use on high school students across all three AI modes defined. From a neuroscience perspective, this highlights the importance of examining not only immediate neural responses but also residual effects that shape ongoing cognitive function. From an engineering perspective, this shapes how AI models are developed to mitigate carryover effects during interaction.

## XVIII. FIELD IMPACTS

This study develops a preliminary framework for examining artificial intelligence interaction as a set of distinct cognitive interaction modes rather than the binary AI versus no-AI conditions. While most existing research evaluates AI exposure as a single construct, the present findings suggest that Tutor, Collaborator, and Solver interactions may be associated with different behavioral profiles and descriptive electrophysiological patterns.

Although no statistically significant global EEG differences were observed across modes, the behavioral findings and mode-specific oscillatory trends indicate that future investigations may benefit from treating AI interaction style as an independent experimental variable rather than aggregating all AI use into a single category.

Beyond the specific findings reported here, the study introduces a methodology that combines controlled AI prompting, behavioral observation, and EEG monitoring within an educational context. This framework may provide a foundation for future research seeking to characterize how different forms of human-AI interaction relate to engagement, problem-solving behavior, and learning processes.

The behavioral metrics developed in this study, including Initiation, Processing, and Stress, may also serve as candidate variables for future adaptive educational systems. Rather than assuming a fixed interaction strategy, future systems could investigate whether AI assistance can be dynamically adjusted according to observed user behavior. However, evaluating the effectiveness of such adaptive approaches will require further experimental validation.

Finally, the results highlight the importance of examining not only immediate responses to AI interaction but also potential short-term carryover effects. While the present findings should be interpreted solely as exploratory, they suggest that temporal dynamics may represent an additional dimension of human-AI interaction worthy of future investigation.

Taken together, this work provides an initial experimental framework for studying mode-dependent AI interaction in high school learners. By suggesting that AI's cognitive impact is mode-dependent and has the potential to create changes at both the neural and behavioral levels, this study lays the foundation for future research aimed at optimizing student–AI collaboration for sustained critical thinking and focus.

## XIX. FUTURE WORK

As this study explores a largely new concept in electroencephalography, there are three key facets that this research should be improved upon in the future.

1. The first is to test a larger demographic of people to expand generalizability. While restricting the sample to bilingual individuals with a 4.0+ weighted GPA enabled tighter control over neurolinguistic and educational variability, it also limited the extent to which findings can be extrapolated to broader student populations. Proper implications of these findings worldwide would require reproducing the methodology across students of all learning levels to determine whether the observed neural and behavioral patterns remain consistent or exhibit population-specific and culture-dependent variation.

2. The second is to examine long-term exposure in participants through a longitudinal study. The 30-minute session performed for all 48 participants illuminated distinct short-term effects across each mode. However, another aspect one must not forget is tracing how the correct usage of AI modes affects students over multiple years. Understanding these trajectories is essential for determining whether the AI modes identified serve as a scaffold for sustaining long-term cognition or eventually end up showcasing different effects in long-term cognitive capacity.

3. A third direction for future testing, and a particularly interesting one, is to test children with differing attention patterns, such as children with ADHD (Attention-Deficit/Hyperactivity Disorder) and ASD (Autism Spectrum Disorder). These groups often exhibit differences in attentional regulation and executive functioning, which may meaningfully alter AI interaction across modes at both behavioral and neural levels. Testing such populations would not only improve inclusivity but also provide insight into whether the unique oscillatory patterns resulting from the study can be adapted to support neurodivergent learners.

## XX. ACKNOWLEDGEMENT

K. Khurana and A. Liew would like to thank Dr. Yu Meng and Professor Shouyi Wang for their review of our research methodology and guidance with scientifically interpreting study conclusions.

## XXI. REFERENCES

[1] S. Wang, X. Tao, H. Ma, F. Li, and C. Wu, "EEG assessment of artificial intelligence-generated content impact on student creative performance and neurophysiological states in product design," Front. Psychol., vol. 16, 2025, doi: 10.3389/fpsyg.2025.1508383.

[2] N. Kosmyna et al., "Project publications: Your brain on ChatGPT," MIT Media Lab, 2025. [Online]. Available: http://www.media.mit.edu/projects/your-brain-on-chatgpt/publications/

[3] I. C. Peláez-Sánchez, D. Velarde-Camaqui, and L. D. Glasserman-Morales, "The impact of large language models on higher education: Exploring the connection between AI and education 4.0," Front. Educ., vol. 9, 2024, doi: 10.3389/feduc.2024.1392091.

[4] V. Jurcak, D. Tsuzuki, and I. Dan, "10–20, 10–10, and 10–5 systems revisited: Their validity as estimates of regional cerebral locations," NeuroImage, vol. 34, no. 4, pp. 1600–1611, Feb. 2007.

[5] E. K. S. Louis, L. C. Frey, et al., "The Normal EEG," American Epilepsy Society, 2016. [Online]. Available: https://www.ncbi.nlm.nih.gov/books/NBK390343/

[6] A. Widmann, E. Schröger, and B. Maess, "Digital filter design for electrophysiological data – a practical approach," J. Neurosci. Methods, vol. 250, pp. 34–46, Sep. 2015.

[7] M. X. Cohen, Analyzing Neural Time Series Data: Theory and Practice. Cambridge, MA, USA: MIT Press, 2014.

[8] E. K. S. Louis et al., "Appendix 3: Principles of electrical safety," in The Normal EEG. American Epilepsy Society, 2016. [Online]. Available: https://www.ncbi.nlm.nih.gov/books/NBK390355/

[9] MNE-Python, "Overview of MEG/EEG analysis with MNE-Python." [Online]. Available: https://mne.tools/stable/auto_Tutorials/intro/10_overview.html

[10] C. S. Nayak and A. C. Anilkumar, "EEG normal waveforms," StatPearls, Mar. 24, 2019. [Online]. Available: https://www.ncbi.nlm.nih.gov/books/NBK539805/

[11] W. Klimesch, "EEG alpha and theta oscillations reflect cognitive and memory performance: A review and analysis," Brain Research Reviews, vol. 29, no. 2–3, pp. 169–195, Apr. 1999.

[12] W. J. Ray and H. W. Cole, "EEG alpha and beta activity during cognitive tasks and component states," Science, vol. 228, no. 4700, pp. 750–752, May 1985.

[13] E. Tan et al., "Theta activity and cognitive functioning: Integrating evidence from task-related developmental EEG research," Dev. Cogn. Neurosci., vol. 67, p. 101404, Jun. 2024.

[14] C. S. Herrmann, H. Strüber, D. Helfrich, and A. K. Engel, "EEG gamma oscillations: From Kilohertz-activity to cognitive magnitudes," Cortex, vol. 62, pp. 25–40, Jan. 2015.

[15] R. T. Canolty and R. T. Knight, "The functional role of cross-frequency coupling," Trends Cogn. Sci., vol. 14, no. 11, pp. 506–515, Nov. 2010.

[16] D. Hagemann and E. Naumann, "The effects of ocular artifacts on (lateralized) broadband power in the EEG," Clin. Neurophysiol., vol. 112, no. 2, pp. 215–231, Feb. 2001.

[17] J. Cohen, "Statistical Power Analysis for the Behavioral Sciences," 2nd ed. Lawrence Erlbaum Associates, 1988.

[18] A. M. Pendry, J. M. Vandagriff, C. H. Y. Wang, K. A. Gee, and E. K. Carr, "Behavior Coding of Adolescent and Therapy Dog Interactions During a Social Stress Task," Veterinary Sciences, vol. 11, no. 12, p. 644, 2024.

[19] M. Friedman, "The use of ranks to avoid the assumption of normality implicit in the analysis of variance," Journal of the American Statistical Association, vol. 32, no. 200, pp. 675–701, Dec. 1937.

[20] F. Wilcoxon, "Individual comparisons by ranking methods," Biometrics Bulletin, vol. 1, no. 6, pp. 80–83, Dec. 1945.

[21] J. Sweller, "Cognitive load during problem solving: Effects on learning," Cogn. Sci., vol. 12, no. 2, pp. 257–285, Apr. 1988.

[22] A. T. Pope, E. H. Bogart, and D. S. Bartolome, "Biocybernetic system evaluates indices of operator engagement in automated task," Biol. Psychol., vol. 40, no. 1, pp. 187–195, 1995.

[23] E. F. Risko and S. J. Gilbert, "Cognitive offloading," Trends in Cogn. Sci., vol. 20, no. 9, pp. 676–688, Sep. 2016.

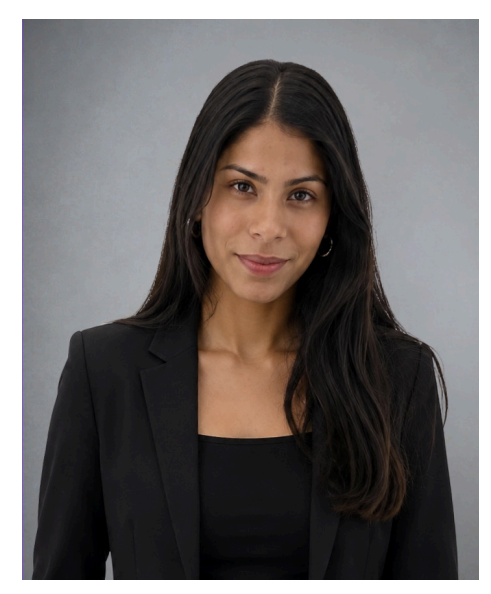

**KASHIKA KHURANA** was born in Panchkula, India, in 2009. She will receive her honors high school diploma through Allen High School in Allen, TX, USA, in May 2027.

Kashika's research experience is centered around using technology and artificial intelligence to aid human health. She has conducted neuroscience, educational, and gut-brain axis research, all incorporating statistical and computational approaches to accelerate intervention.

Khurana has presented her work at multiple competitive venues, including the Dallas Science and Engineering Fair (2025, 2026), the Texas Science and Engineering Fair (2026), and was selected for the MIT App Inventor Global AI Summit (2026), where her projects were recognized for their innovation and interdisciplinary approach, combining behavioral health with rigorous engineering methodology.

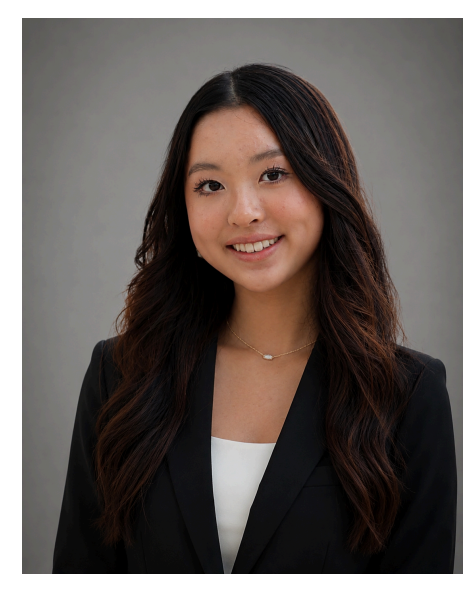

**ALLY LIEW** was born in Irving, TX, USA, in 2008. She is pursuing her high school education at Allen High School, Allen, TX, USA, and is expected to graduate in May 2027. Her academic focus includes science research and interdisciplinary studies.

She has participated in independent research and competitive science programs. In March 2026, she was recognized as a finalist at the Texas Science and Engineering Fair, and was selected for the MIT App Inventor Global AI Summit in May 2026. Her current research interests include neuroscience, behavioral analysis, and interdisciplinary problem-solving.

Liew continues to engage in academic and research-oriented activities and plans to pursue advanced study in a STEM-related field.